 \documentclass[final,5p,times,twocolumn]
 {elsarticle}
 
\usepackage{amsmath}
\usepackage{amssymb}
\usepackage{lipsum}
\usepackage{pdfpages}
\usepackage{afterpage}

\newcommand\blankpage{%
    \null
    \thispagestyle{empty}%
    \addtocounter{page}{-1}%
    \newpage}
    
\usepackage[pdftex,colorlinks,linkcolor = blue,urlcolor  = blue,citecolor = blue]{hyperref}

\newcommand{\kms}{km\,s$^{-1}$}

\journal{Physics Letters B}

\begin{document}

\begin{frontmatter}


\title{Systematics of the low-energy electric dipole strength in the Sn isotopic chain}


\author[a]{M. Markova}\ead{maria.markova@fys.uio.no}
\affiliation[a]{organization={Department of Physics, University of Oslo},
            city={N-0316 Oslo},
            country={Norway}}
            


\author[b]{P. von Neumann-Cosel}\ead{vnc@ikp.tu-darmstadt.de}
\affiliation[b]{organization={Institut für Kernphysik, Technische Universität Darmstadt, D-64289 Darmstadt, Germany},
           city={D-64289 Darmstadt},
           country={Germany}}
            

\author[c,d,e]{E.~Litvinova}\ead{elena.litvinova@wmich.edu}
\affiliation[c]{organization={Department of Physics, Western Michigan University},
            city={Kalamazoo},
            postcode={Michigan 49008},
            country={USA}}
\affiliation[d]{organization={National Superconducting Cyclotron Laboratory, Michigan State University},
            city={East Lansing},
            postcode={Michigan 48824},
             country={USA}}
\affiliation[e]{organization={GANIL, CEA/DRF-CNRS/IN2P3},
            city={F-14076 Caen},
            country={France}}
            
\begin{abstract}
We present a systematic study of the mass dependence of the low-energy electric dipole strength (LEDS) in Sn isotopes in the range $ A = 111 - 124$ based on data obtained with the Oslo method and with relativistic Coulomb excitation in forward-angle ($p,p^\prime$) scattering.
The combined data cover an energy range of $2 - 20$ MeV which permits, with minimal assumptions, a decomposition of the total strength into the contribution from the low-energy tail of the isovector giant dipole resonance (IVGDR) and possible resonance-like structures on top of it.
In all cases, a resonance peaked at about 8.3 MeV is observed, exhausting an approximately constant fraction of the Thomas-Reiche-Kuhn (TRK) sum rule.
For heavier isotopes ($A \geq 118$) a consistent description of the data requires the inclusion of a second resonance centered at 6.5 MeV, representing the isovector response of the pygmy dipole resonance (PDR). 
Its strength corresponds to a small fraction of the total LEDS only and shows an approximately linear dependence on mass number.
The experimental results are compared to microscopic calculations to investigate the importance of an inclusion of quasiparticle vibration coupling (qPVC) for a realistic description of the LEDS.
A possible interpretation of the experimentally observed two-bump structure is given.
\end{abstract}




\end{frontmatter}




\section{Introduction}
\label{introduction}

The observation of a resonance-like structure in the electric dipole response of heavy nuclei at energies around or below the neutron threshold, commonly termed pygmy dipole resonance (PDR), has been a topic leading to considerable experimental and theoretical activities in recent years \cite{Paar2007,Savran2013,Bracco2019,Lanza2022}.
The interest has been triggered by attempts to interpret the underlying structure and also investigate its impact on the cross sections of ($n,\gamma$) reactions relevant to a deeper understanding of the nucleosynthesis of heavy elements \cite{Goriely2023}.
Qualitatively, all mean-field-based models predict the appearance of such a mode, however, with a broad range of strengths and energy centroids depending on a chosen interaction, a model space, and possible extensions beyond the random phase approximation (RPA) level. 

Many theoretical studies interpret the PDR to arise from neutron skin oscillations, which implies a dependence of the PDR strength on neutron excess \cite{Piekarewicz2006,Klimkiewicz2007,Tsoneva2008,Carbone2010,Piekarewicz2011,Ebata2014,Baran2014}. 
Accordingly, some models predict a correlation of the PDR strength with neutron skin thickness, which in turn suggests that information on the symmetry energy parameters of neutron-rich matter can be derived (see e.g.\ \cite{Klimkiewicz2007,Carbone2010}).
These claims, however, have been questioned \cite{Litvinova2009,Reinhard2010,Urban2012,Reinhard2013,Papakonstantinou2014}.
The neutron skin oscillation interpretation has been mainly based on a specific form of transition densities in the energy region of the PDR with an approximately isoscalar (IS) radial dependence in the interior and a peak of the neutron density at the surface.
The dominance of the neutron character of the $E1$ transitions forming the PDR is experimentally confirmed by studies of the IS response with $(\alpha,\alpha^\prime\gamma$) \cite{Savran2006,Endres2010} and ($^{17}{\rm O},^{17}{\rm O}^\prime\gamma)$ \cite{Crespi2014,Pellegri2014} reactions.
However, all experimental signatures of the PDR, viz., large ground-state branching ratios, large isovector (IV) strengths (with respect to average B(E1) transition strengths at low excitation energies), significant IS strengths, as well as the characteristic form of the transition density in heavy nuclei, are consistent with an interpretation as a low-energy IS toroidal mode \cite{Repko2019}.
The recent first experimental demonstration \cite{Neumann-Cosel2023} of the toroidal nature of low-energy $E1$ transitions with all the experimental signatures quoted above, albeit in a $N \approx Z$ nucleus, and successful description of these data by models predicting a dominantly toroidal nature of the PDR further challenge the neutron skin oscillation picture. 

The best suited case for a systematic investigation of a dependence of PDR properties on neutron excess is the Sn isotope chain between $A = 100$ and 132. Despite covering a variation from neutron shell closure to mid-shell, the proton shell closure stabilizes the ground state (g.s.) features leading to similar low-energy structure along the chain.
Thus, theoretical work has been focused on the Sn chain, attempting to predict the evolution of the PDR photoabsorption cross sections with neutron excess in order to test a possible relation to the neutron skin thickness \cite{Piekarewicz2006,Klimkiewicz2007,Tsoneva2008,Carbone2010,Piekarewicz2011,Ebata2014,Baran2014,Litvinova2009,Papakonstantinou2014}.
However, many of these studies use a summed strength below some cutoff energy to represent the PDR photoabsorption cross sections, despite this approach not being supported by the data.
Experimental studies of the IS component of the PDR have demonstrated an isospin splitting (for $^{124}$Sn see \cite{Endres2010,Pellegri2014,Endres2012}), where good correspondence between transitions excited in the IS and IV response is observed in a confined energy region at lower excitation energies, while at higher excitation energies the IV response dominates.

Furthermore, a comparison of photoabsorption cross sections in $^{112,116,120,124}$Sn \cite{Bassauer2020b,Krumbholz2015} deduced from Coulomb excitation in forward-angle $(p,p^\prime)$ scattering \cite{Neumann-Cosel2019a} and from the $(\gamma,\gamma^\prime)$ reaction \cite{Ozel-Tashenov2014,Govaert98} shows good agreement in the energy region where the IS strength was also found, but large differences at higher excitation energies.
The strengths deduced from the $(\gamma,\gamma^\prime)$ experiments depend on the g.s.\ branching ratio \cite{Zilges2022}.
The dramatic reduction compared to the $(p,p^\prime)$ results points to a complex structure of the excited states with small g.s.\ branching ratios, suggesting that they rather belong to the low-energy tail of the IVGDR.
Thus, a significant part of the low-energy photoabsorption strength is {\it not} related to the PDR.
This interpretation has recently been confirmed by a new $^{120}$Sn$(\gamma,\gamma^\prime)$ experiment \cite{Muscher2020}, where the contributions from the statistical decay were additionally extracted. 

Here we present a systematic study of the LEDS in Sn isotopes with mass numbers from 111 to 124. 
It is based on a new set of Gamma Strength Functions (GSFs) from $\gamma$ decay after light-ion-induced compound reactions (the so-called Oslo method) \cite{Markova2021,Markova2022,Markova2023}  combined with a recent study of stable even-mass isotopes with relativistic Coulomb excitation in the $(p,p^\prime)$ reaction \cite{Bassauer2020b,Bassauer2020a}. 
The GSFs are directly related to the photoabsorption cross sections, provided the Brink-Axel hypothesis \cite{Brink1955,Axel1962} also holds for the LEDS.
This seems to be the case for Sn isotopes \cite{Markova2021,Markova2022,Markova2023}.
The combined data sets allow a decomposition of the LEDS with minimal assumptions, providing information on the evolution of the PDR strength and other components with neutron excess.
The extraction of the GSFs from the Oslo data and a critical investigation of the assumptions underlying the decomposition and possible variants thereof are discussed in detail in Ref.~\cite{Markova2024}. 
The experimental GSFs are then compared to microscopic calculations to investigate the importance of an inclusion of quasiparticle vibration coupling (qPVC) for a description of the LEDS in order to provide an interpretation of the apparent structures.

\section{Experimental data}

GSFs extracted with the Oslo method are available for $^{111-113,116-122,124}$Sn.
They typically cover a $\gamma$ energy range from about 2 MeV to $1 -2 $ MeV below the respective neutron thresholds ($S_n$). 
The results for $^{111-113}$Sn, $^{116}$Sn, and $^{118}$Sn  were obtained with a custom-designed Si telescope ring (SiRi) \cite{Guttormsen2011} combined with the CACTUS $\gamma$ ball \cite{Guttormsen1990_CACTUS} made of 28 5$^{\prime\prime}\times5^{\prime\prime}$ NaI detectors.
New experiments were performed for $^{117,119,120,124}$Sn using the OSCAR detector system \cite{Ingeberg2020,Zeiser2020}, consisting of 30 large-volume LaBr $\gamma$ detectors providing superior efficiency, timing, and energy resolution compared to the previous experimental setup.
Therefore, the new data for $^{117,119}$Sn supersede previous results.  
Details of the consistent extraction of nuclear level densities and GSFs are described in Ref.~\cite{Markova2024}. 

An independent set of GSFs for the even-mass isotopes $^{112-120,124}$Sn is available from relativistic Coulomb excitation in the $(p,p^\prime)$ reaction at extreme forward angles \cite{Bassauer2020b}. 
They agree in all cases with the Oslo method results within the respective error bars. 
This implies that the Brink-Axel hypothesis underlying the Oslo method and controversially discussed for the LEDS (see e.g.\ Refs.~\cite{Netterdon2015,Guttormsen2016,Martin2017,Isaak2019}) holds in case of the Sn isotopes.
Data from the latter experiment are available in an energy range $6 - 20$ MeV, covering the major part of the IVGDR and providing sufficient overlap with the Oslo method results.
It should be noted that previous photoabsorption experiments using the $(\gamma,xn)$ reaction show considerable scatter in the systematics of IVGDR parameters. 
In contrast, the results of Ref.~\cite{Bassauer2020b} provide centroid energies in line with empirical systematics and an almost constant width, as expected from the similar g.s.\ structure.
In the energy region close to $S_n$, where discrepancies with older data are particularly pronounced, they are also in good agreement with new ($\gamma,n$) experiments \cite{Utsunomiya2009,Utsunomiya2011} using monoenergetic photon beams from laser Compton back-scattering. 

\section{Decomposition of the low-energy electric dipole strength}
\label{section:decomposition}


\begin{figure}[t]
\includegraphics[width=0.965\columnwidth]{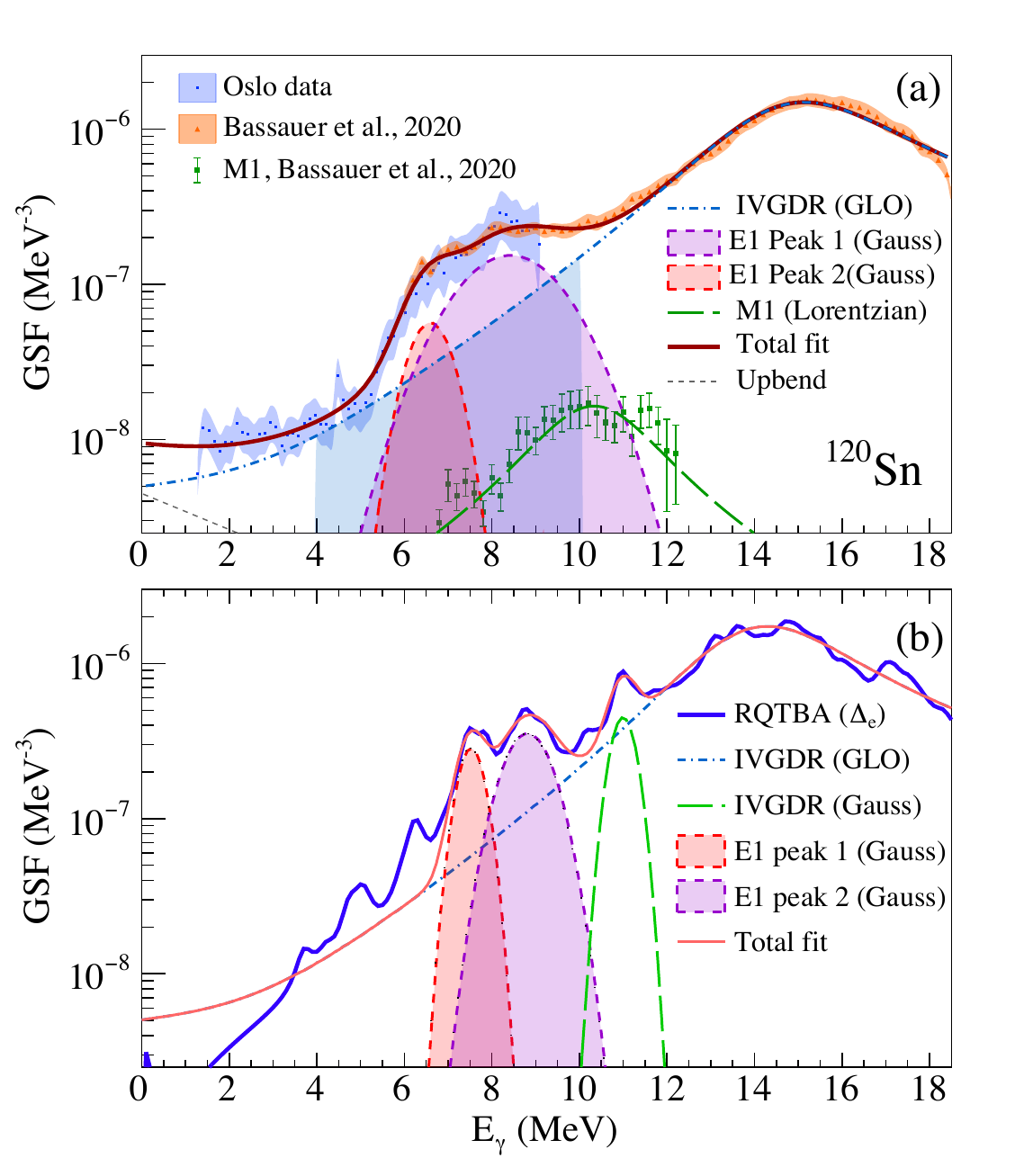}
\caption{\label{fig1}
(a) Decomposition of the experimental GSF of $^{120}$Sn. The Oslo data (blue band) are  shown together with the $(p,p^{\prime})$ results \cite{Bassauer2020b} (orange band). 
The fit of the IVGDR with the GLO model \cite{Capote2009}, the total fit, and the low-lying $E1$ components are shown in (a) as blue dash-dotted line, red solid line, and red and violet shaded areas, respectively. 
The $M1$ data from the (p,p$^\prime$) experiment \cite{Bassauer2020b} and the corresponding Lorentzian fit are shown as green data points and dashed line, and the upbend at low $\gamma$ energies as grey dashed line. (b) Same as (a) for the RQTBA results.
An additional Gaussian (green dashed line) is added to reproduce the IVGDR energy region.}
\end{figure}

A decomposition of the LEDS  performed for the example of $^{120}$Sn is illustrated in Fig.~\ref{fig1}(a).
The following contributions to the total GSF are considered: a low-energy tail of the IVGDR, a spin-flip M1 resonance, an upbend at very low energies and one or two (when demanded by the data) E1 resonances.
The generalized Lorentzian model \cite{Capote2009} (solid blue line) is chosen to describe the IVGDR part.
%
%
It is the only empirical or microscopic model capable of accounting simultaneously for the low-energy flank of the IVGDR and a relatively flat strength distribution at very low energies ($2 - 4$ MeV). 

Data on the M1 spin-flip resonance are available (green points) in the energy region from 6 to about 12 MeV from Ref.~\cite{Bassauer2020b} and show broad distributions. 
They are parameterized by a single Lorentzian (dashed green line). 
The upbend at very low energies is described by exponential functions (dashed gray line).
The determination of parameters is described in Ref.~\cite{Markova2024}, but its details have no influence on the present results. 
Finally, additional resonances on top of the tail of the IVGDR are assumed to be Gaussian (shaded red and violet areas).
All parameters are determined by a simultaneous fit (dark red line) to the combined Oslo (blue points and error band) and $(p,p^\prime)$ (orange points and error band) data described above.
An in-depth discussion of the decomposition is provided in Ref.~\cite{Markova2024}.
A corresponding fit of the theoretical strength distribution shown in Fig.~\ref{fig1}(b) is discussed in the next Section.

\begin{figure}
\includegraphics[width=0.98\columnwidth]{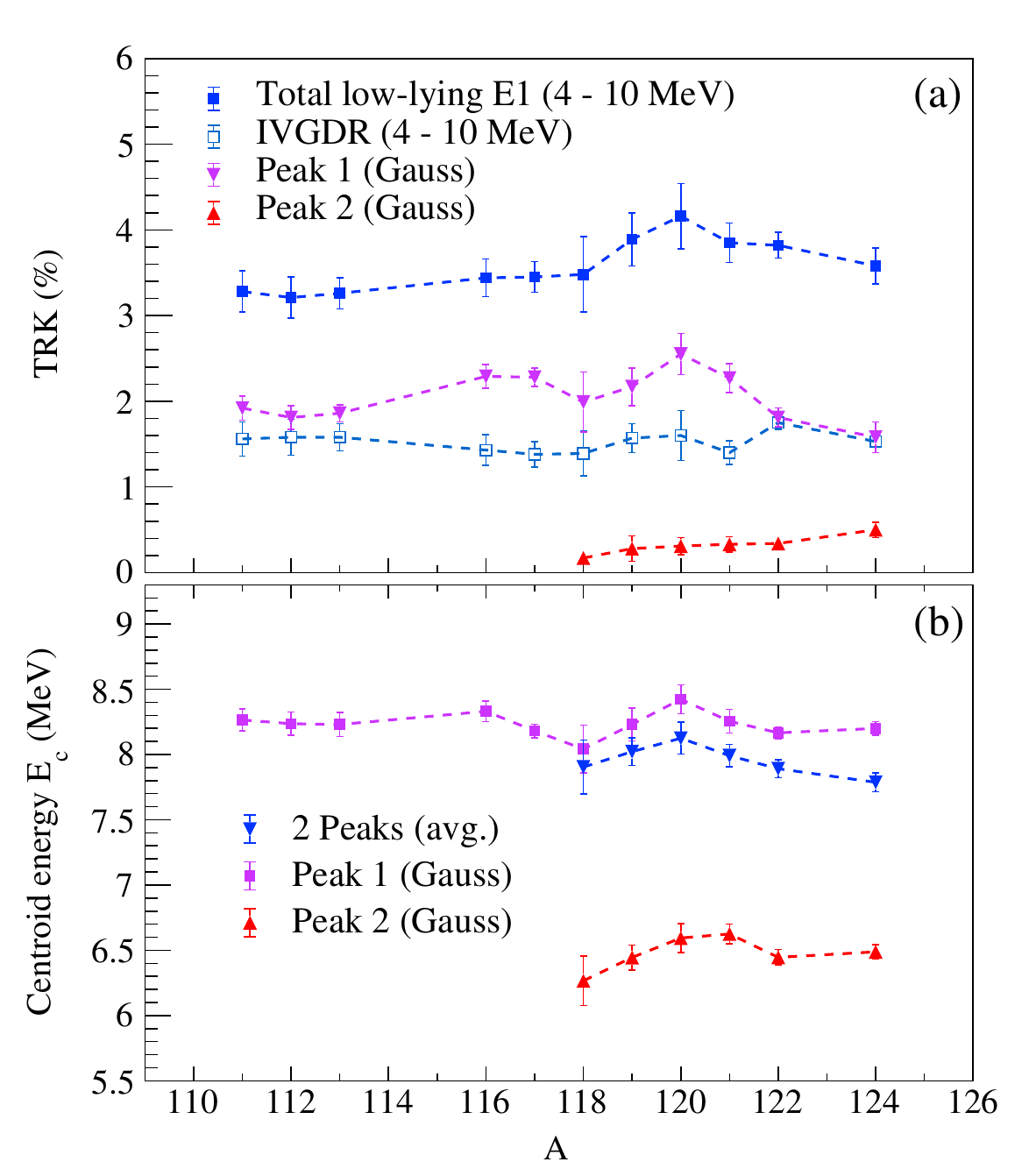}
\caption{\label{fig2}
Systematics of the total LEDS  integrated over the energy region $4 -10$ MeV and its decomposition into the contributions from the tail of the IVGDR and one or two (for masses $\geq 118$) resonances on top.
(a) Strengths in \% of the TRK sum rule.
(b) centroid energies.}
\end{figure}

Figure \ref{fig2} presents the evolution of the LEDS and its various components with mass number when integrating within the $4 - 10$ MeV energy region.
The total LEDS is roughly constant (with slight indications of a local maximum at $^{120}$Sn) with values ranging from 3.5\% to 4\% of the classical Thomas-Reiche-Kuhn (TRK) sum rule.
The contribution of the low-energy tail of the IVGDR is found to be approximately constant exhausting a TRK value of about 1.5\%.
The strength on top of the tail of the IVGDR in the isotopes with $A \leq 118$ is well parameterized by a single Gaussian function.
For $A \geq 118$ the $(p,p^\prime)$ data demand the inclusion of a second peak at lower energies, which is particularly pronounced in $^{124}$Sn, cf.\ Fig.~\ref{fig1}(a).
The dominant peak is centered at about 8.3 MeV and its average strength is 2\% of the TRK value.
Small variations are visible as a function of mass number with a possible strength maximum at $^{120}$Sn. 
A similar local maximum of the LEDS at $^{120}$Sn has been discussed e.g.\ in Refs.~\cite{Paar2007,Piekarewicz2006,Burrello2019}.

The centroid of the lower-energy peak lies at about 6.5 MeV with no energy dependence within uncertainties.
Its strength is small, ranging from 0.1\% for $^{118}$Sn to a maximum of 0.5\% of the TRK sum rule in $^{124}$Sn. 
Interestingly, this contribution shows an approximately linear dependence on mass number (or neutron excess).
When extrapolated to lower masses, its predicted strength is too small to be distinguished from the broad peak at higher energies.
Studies of the isoscalar E1 strength in $^{124}$Sn with the  $(\alpha,\alpha^\prime\gamma$) \cite{Endres2010,Endres2012} and ($^{17}{\rm O},^{17}{\rm O}^\prime\gamma)$ \cite{Pellegri2014} reactions show a concentration between 5.5 and 7 MeV, consistent with the properties of the lower-energy peak.
Furthermore, studies of the $^{112,116,120,124}$Sn$(\gamma,\gamma^\prime)$ reaction \cite{Ozel-Tashenov2014,Govaert98} typically find the strongest transitions between 6 and 7 MeV, indicating large g.s.\ branching ratios.
These experimental signatures point towards an interpretation of the lower-energy peak as the IV response of the PDR.
Similar arguments have been put forward for $^{120}$Sn based on a detailed analysis of the particle-hole structure of $1^-$ states excited in the $^{119}$Sn($d,p\gamma$) reaction \cite{Weinert2021}.

These numbers are significantly smaller than quoted in most theoretical studies of the dependence of the PDR strength on neutron excess in the Sn isotope chain (for a notable exception see Ref.~\cite{Tsoneva2008}).
They challenge an interpretation of the PDR as neutron skin oscillation, which implies some degree of collectivity.
Of course, the distinction of strength belonging either to the PDR or to the tail of the IVGDR is schematic.
In reality, there is some degree of mixing between transitions of different types.
All QRPA calculations agree on a gradual change of the radial transition densities from  an approximate IS behavior in the interior of the nucleus and a pronounced peak of the neutron density at the surface (a signature of the PDR) to more IV-type transitions at higher energies. 
However, for the confined energy region where the PDR strength is located, effects on the IV response due to variations of the transition densities are expected to be small.

The discrepancy probably arises in many cases from the fact that QRPA results typically produce a single low-energy peak and its strength is assumed to represent the PDR strength. 
The present work and experiments comparing the IS and IV response on the same nucleus demonstrate that this is {\it not} correct.
The PDR strength exhausts a rather small fraction of the LEDS only (at most about 15\% for the Sn isotopes studied here).
Thus, quantitative predictions based on QRPA have to be taken with some care, and one might have to go beyond it by including complex configurations to achieve realistic low-energy strength distributions.
This is discussed in the next section.

\section{Comparison with microscopic calculations}


In the following, we compare the experimental GSFs with microscopic calculations using $^{120}$Sn as a representative example.
Details of the model and a comparison with all experimental results can be found in Ref.~\cite{Markova2024}.
Nuclear response theory can be consistently derived in the model-independent ab initio equation of motion (EOM) framework \cite{AdachiSchuck1989, DukelskyRoepkeSchuck1998,LitvinovaSchuck2019} for the in-medium two-time two-fermion propagators. In superfluid media, particle-hole ($ph$), particle-particle ($pp$), and hole-hole ($hh$) propagators can be conveniently unified in one two-quasiparticle ($2q$) propagator without loss of generality \cite{Litvinova2022}.

While the generic EOM for the correlated (four-time) two-fermion propagator is the Bethe-Salpeter equation, the two-time character of the response leads to the EOM of a Dyson form, which depends on a single time difference.  The interaction kernel of the Bethe-Salpeter-Dyson equation, before taking any approximation, decomposes into the static and dynamical (time-dependent) components. In the energy domain, they are represented by the energy-independent and energy-dependent terms, respectively. The energy-dependent contribution generates long-range correlations while making an impact on their short-range static counterpart. Since the $2q$-propagator EOM couples to a growing hierarchy of higher-rank EOMs via the dynamical kernel, in practical applications, it is decoupled by making approximations with varying correlation content. 

The simplest decoupling scheme retains only the static kernel and is known as the quasiparticle random phase approximation (QRPA). The heart of the dynamical kernel is the four-fermion $4q$ fully correlated propagator contracted with two interaction matrix elements. The approximation keeping the leading effects of emergent collectivity in the $4q$ propagator reduces it to $2q\otimes phonon$ configurations, where the $phonon$ represents correlated $2q$ pairs (vibrations). This approach reproduces the phenomenological (relativistic) quasiparticle time-blocking approximation ((R)QTBA) \cite{Tselyaev2007,LitvinovaRingTselyaev2008} if the bare interaction between nucleons is replaced by their effective interaction. However, in the ab initio EOM method, the quasiparticle-vibration coupling (qPVC) amplitudes are derived consistently from the underlying interaction, while in (R)QTBA the qPVC self-energy is used as an input. The EOM works with the two-time propagators from the beginning, in contrast to QTBA operating four-time propagators, and hence does not employ time blocking. Furthermore, the ab initio relativistic  EOM (REOM) framework links the nuclear response theory to the underlying scale of particle physics and fundamental interactions, in the present case, to the meson-exchange interaction and, more importantly, enables extensions of the $4q$ dynamical kernel to more complex configurations \cite{Litvinova2022}. 

The cluster decomposition of this kernel identifies the next-level complexity non-perturbative approximation as the $2q\otimes 2phonon$. The implementation of such configurations is becoming gradually possible with the increasing computational capabilities \cite{LitvinovaSchuck2019,Litvinova2023a} and will be applied for systematic calculations in the near future. While implementations with bare interactions are not yet available, (R)EOM admits realistic implementations that employ effective interactions adjusted in the framework of the density functional theory. For such interactions, the qPVC in the dynamical kernel can be combined with subtraction, restoring the self-consistency of the framework \cite{Tselyaev2013}, while reasonable phonons can be obtained already on the QRPA level. 

In this work, the effective meson-exchange interaction NL3*  \cite{NL3star} was employed. This interaction has a transparent link to particle physics, while the meson masses and coupling constants are only slightly different from their vacuum values (cf., for instance, Ref. \cite{Machleidt1989}). The NL3* is an upgraded version of the previously developed NL3 parametrization \cite{Lalazissis1997}, which includes self-interactions in the scalar $\sigma$ meson sector \cite{BogutaBodmer1977}. The self-interactions represent three and four $\sigma$ meson vertices, i.e., concrete physical processes that occur in correlated media. The NL3 and NL3* ans\"atze are separable in the momentum representation, which speeds up the computation considerably. With these interactions, the realization of the REOM approach on the $2q\otimes phonon$ level technically corresponds to RQTBA, so this name is retained in the present work.

\begin{figure}[t]
\includegraphics[width=\columnwidth]{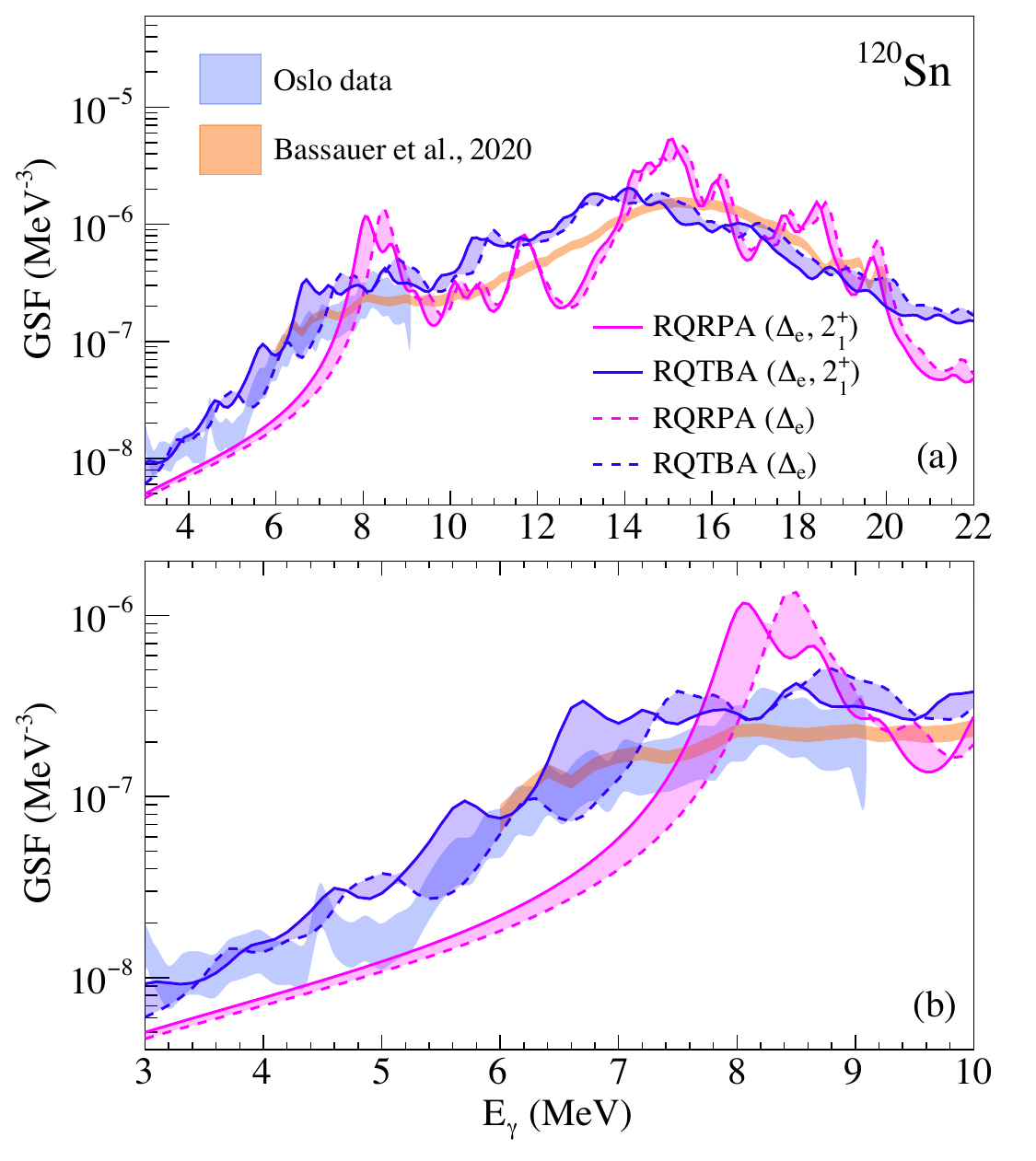}
\caption{\label{fig3}
Comparison of experimental and calculated GSFs computed with an artificial width of 200 keV for $^{120}$Sn in the energy regions (a) $3 - 22$ MeV and (b) $3 - 10$ MeV.
The blue and orange bands indicate the Oslo-method and $(p,p^{\prime})$ data. 
The RQRPA and RQTBA calculations described in the text are shown as magenta and violet lines, respectively.}
\end{figure}

Results of the RQRPA and RQTBA calculations for the electric dipole strength in $^{120}$Sn are displayed in Fig.~\ref{fig3} together with the data.
The ($p,p^\prime$) results in the energy region of the IVGDR (orange band in Fig.~\ref{fig3}(a)), are reasonably well described by both approaches.
The successful reproduction of the gross features, such as the centroid, total strength, and Landau damping, by the RQRPA calculations highlights the prime importance of the high-energy $2q$ configurations for the IVGDR formation in heavy nuclei \cite{Neumann-Cosel2019b}. 

Pairing correlations underly superfluidity in open-shell nuclei, which is another mechanism responsible for their spectral features. For high-frequency oscillation modes, pairing manifests mostly as fine-tuning their centroid, while at lower frequencies, it may affect the spectrum more sizeably.  Since nuclear superfluidity is dominated by monopole Cooper pairs, in the dipole channel, pairing is expressed via the energies of Bogoliubov quasiparticles $E_i = \sqrt{\varepsilon_i^2 + \Delta_i^2}$, where $\varepsilon_i$ are the mean-field nucleonic energies on the orbits $|i\rangle$, counted from the chemical potential, and $\Delta_i$ are the pairing gaps \cite{Suhonen}.
%
The strength of pairing interaction is not well constrained, and there is some freedom in choosing this parameter \cite{AfanasjevLitvinova2015}. Fig.~\ref{fig3} illustrates, in particular, the sensitivity of the strength functions to the values of $\Delta_i$ distinguishing the results obtained with the empirical pairing gap $\Delta_e = 12/\sqrt{A}$ MeV (RQRPA  ($\Delta_e$), RQTBA ($\Delta_e$)), and those with the pairing gap reproducing the experimental position of the lowest quadrupole state within RQRPA (RQRPA ($\Delta_e, 2^+_1$), RQTBA ($\Delta_e, 2^+_1$)). One can see an enhanced sensitivity of the LEDS to the choice of the pairing strength parameter. In the analysis below, we will focus on the latter version of the theory based on its performance in Ref.~\cite{Litvinova2023}.

An expanded view of the low-energy region is given in Fig.~\ref{fig3}(b), where significant differences between the RQRPA and RQTBA results are apparent.
With RQRPA, the LEDS is dominated by the transition at 8.06 MeV, and the strength at lower energies is only an artifact of the 200 keV width used in the calculations.
The inclusion of qPVC provides a satisfactory description of the energy dependence of the LEDS on a quantitative level down to energies of about 3 MeV, within the uncertainty due to the poorly-known pairing strength, showing the splitting of the RQRPA mode into a large number of states.
The differences in the description of the LEDS underline the problems of an interpretation of the PDR based on calculations on the QRPA level.
The strengths predicted by QRPA correspond to the total LEDS rather than the PDR, while the latter carries a fraction only.
Thus, an analysis of the structures underlying the PDR and the dominant IV resonance at higher energies based on mean-field models should include qPVC, at least in the leading approximation.

The encouraging results motivated an attempt to decompose the RQTBA GSF analog to the experimental data as shown in Fig.~\ref{fig1}(b) for the example of $^{120}$Sn.
An extra Gaussian component around 11 MeV is needed to reproduce the IVGDR region. The two-bump structure in the low-energy region is reproduced with energy centroids and strengths comparable to experiment, but somewhat smaller widths. While such a decomposition provides satisfactory agreement for $^{116,118,120}$Sn, the fits in $^{122,124}$Sn are more complex due to large deviations of the RQTBA predictions from a Lorentzian shape of the IVGDR just above the neutron separation threshold and the larger fragmentation of the low-lying strength. All results and a more detailed discussion of the fitting procedure are provided in the Supplementary Material.

\begin{figure}
\includegraphics[width=\columnwidth]{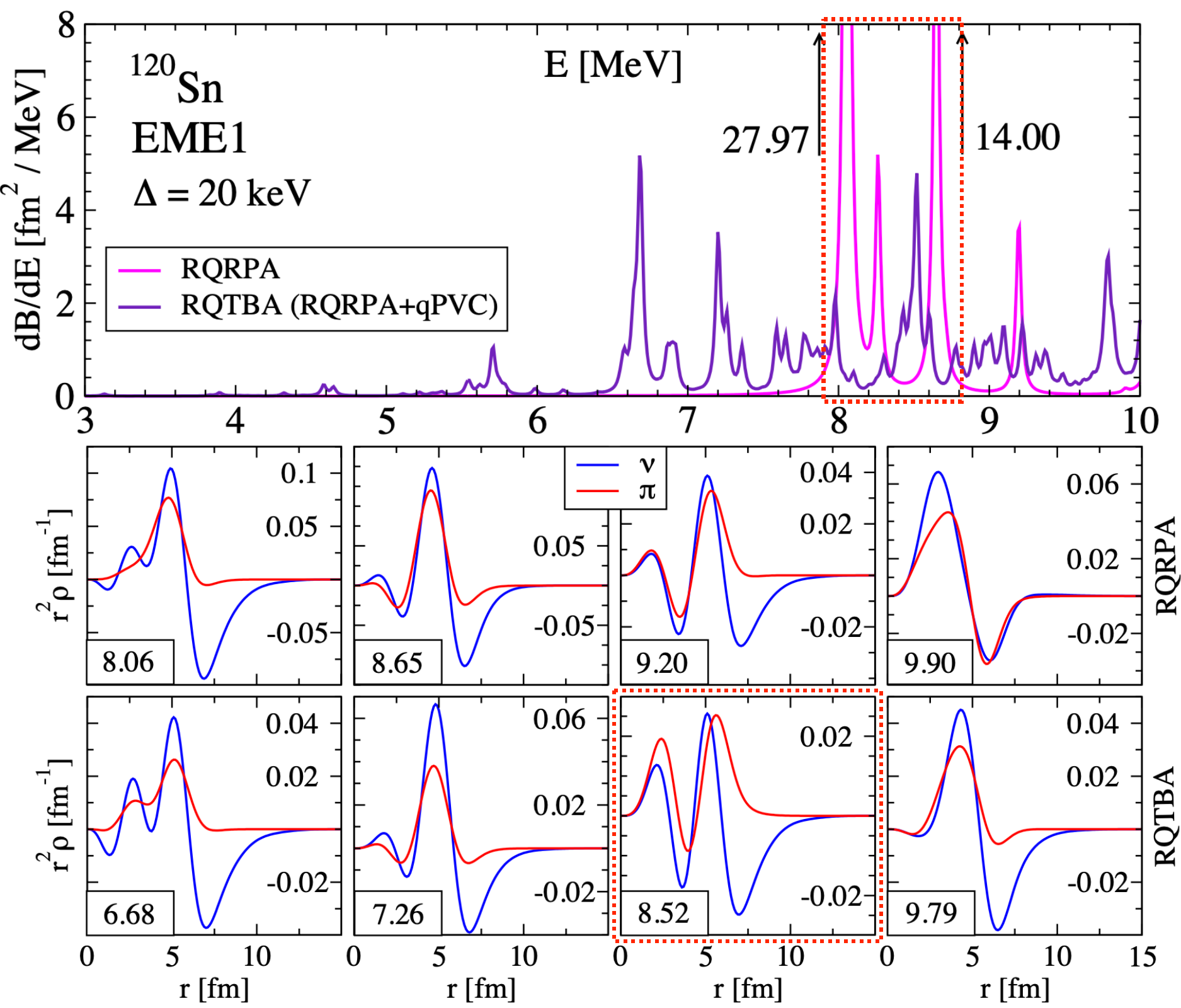}
\caption{\label{fig4}
Top: Theoretical LEDS in $^{120}$Sn with 20 keV smearing. Middle: RQRPA neutron and proton transition densities $\rho^{(\nu,\pi)}(r)$. Bottom: $\rho^{(\nu,\pi)}(r)$ for the RQTBA characteristic states between 6 and 10 MeV. The energies of the states are indicated in the boxes in the bottom-left corners of each panel displaying $\rho^{(\nu,\pi)}(r)$. The dotted box marks the area dominated by the intruder RQTBA states defined in the text.}
\end{figure}
%

To further analyze the nature of the dipole strength, we extract the transition densities of each peak seen in the computation with the 20 keV half-width (smearing parameter) in both approaches in the energy interval $6-10$ MeV.
These spectra are shown in the top panel of Fig. \ref{fig4}. The RQRPA calculations generate two prominent peaks at 8.06 and 8.65 MeV, and two minor peaks at 9.20 and 9.90 MeV
(the latter is visible only in the isoscalar dipole channel). 
The qPVC correlations included in RQTBA yield many more states in this energy region, which we partitioned into 1 MeV energy intervals. For each of them, a characteristic state was chosen in a way that such a state exhibits the maximal $2q$ content in terms of the quantity ${\cal Z}_{ij} = |{\cal X}^n_{ij}|^2 - |{\cal Y}^n_{ij}|^2$, which is the $2q$ contribution to the state normalization and the indices $\{i,j\}$ stand for the quasiparticle orbits if the mean-field basis of Dirac-Hartree-Bogoliubov spinors \cite{LitvinovaRingTselyaev2008}. 
In the (R)EOM formalism, $\{{\cal X}^n_{ij}, {\cal Y}^n_{ij}\} = \rho^n_{ij}$ are the components of the superfluid transition density, such as 
${\cal X}^{n}_{jk} = \langle 0|\alpha_k\alpha_j|n\rangle$ and ${\cal Y}^{n}_{jk} = \langle 0|\alpha^{\dagger}_{j}\alpha^{\dagger}_k|n\rangle$, where
$\{\alpha^{\dagger}_j,\alpha_j\}$ are the operators of the Bogoliubov quasiparticles, and $|0\rangle$ and $|n\rangle$ are the ground and excited states, respectively. Note here that in (R)EOM, the $\{{\cal X}^n_{ij}, {\cal Y}^n_{ij}\}$ amplitudes are of a more general character than those of (R)QRPA because of higher complexity correlations encoded in the structure of the many-body states \cite{Litvinova2022}.

As the leading-order photon field is a single-particle operator, it couples directly to the $2q$ configurations, while the remaining  $2q\otimes phonon$ contributions are manifested through fragmentation of the "primordial" $2q$ configurations of the RQRPA. Thus, the RQTBA states with maximal $\cal Z$-values 
correspond to the
most pronounced peaks. Such states were then chosen as {\it characteristic states} for each 1 MeV energy bin. 
The four RQRPA states automatically qualify as characteristic states for the RQRPA spectrum, which is then naturally partitioned into 0.5 MeV energy intervals. 

While a full account of the RQTBA transition densities is provided in the Supplementary Material, the neutron and proton transition densities $\rho^{(\nu,\pi)}$ of the characteristic states are plotted in the middle and bottom panels of Fig.~\ref{fig4} as functions of the radial distance $r$ from the nuclear center weighted with $r^2$ for RQRPA and RQTBA, respectively. The RQRPA states at 8.06, 8.65, and 9.20 MeV exhibit very similar behavior of $\rho^{(\nu,\pi)}$ with their in-phase oscillation in the bulk and nearly pure neutron oscillation dominance beyond the nuclear surface radius whose mean-field root mean square amounts 4.71 fm. The state at 9.90 MeV is an example of an almost pure isoscalar state that has comparable proton and neutron oscillation contributions canceling out because of the opposite 
sign of the IV effective charges.
The $2q$ components of the RQTBA states generally inherit the structure of the RQRPA primordial states, which is reflected by the $\rho^{(\nu,\pi)}$ of the characteristic states displayed in the bottom panels of Fig. \ref{fig4}. However, one notes immediately the different underlying structure of the RQTBA state at 8.52 MeV, which is representative of the most prominent states located between $\sim$7.9 and $\sim$8.8 MeV. This suggests that this energy region is dominated by "intruder" fragments of the RQRPA states located above 10 MeV, based on the fact that the transition densities around the GDR peak are well established as purely out of phase and the evidence that the GDR's low-energy shoulder originates from $2q$ states of mixed underlying structure interim between the neutron skin oscillation and the GDR-type oscillations. In Ref.~\cite{Endres2010}, it was found that such intruder states are notably suppressed in the isoscalar dipole channel, and in this work, we further dwell on their microscopic structure.

\begin{figure}
\vspace{-1.3cm}
\includegraphics[width=\columnwidth]{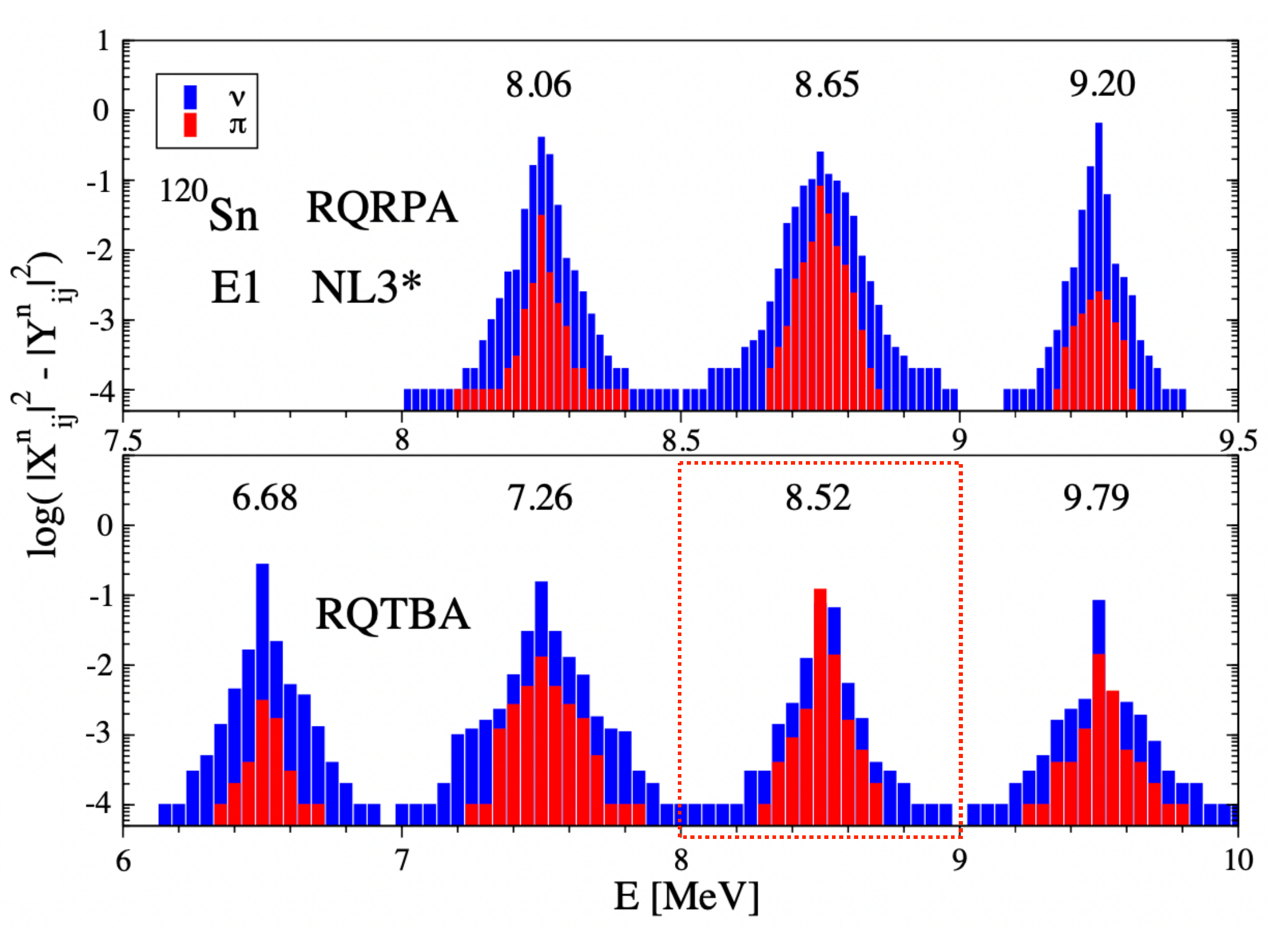}
\caption{\label{fig5}
 $\cal Z$-values exceeding 0.01 \% for characteristic states of RQRPA (top) and RQTBA (bottom). See text for details.
 Proton and neutron amplitudes are plotted in red and blue, respectively.}
\end{figure}

Namely, we collected the $\cal Z$-values of the characteristic states and plotted those exceeding 0.01 \% in Fig. \ref{fig5} placing the largest $\cal Z$-values ${\cal Z}_{ij}^{max}$ in the center of each energy interval and the other ones right and left from ${\cal Z}_{ij}^{max}$ in descending order. One can observe that the intruder states represented by the characteristic state at 8.52 MeV are associated not only with different radial behavior of their transition densities but also with the pronounced dominance of the proton component in their $2q$ composition. 
This suggests that such states are more likely to decay into the excited states than the neutron-dominated ones, which preferentially decay to the ground state.  
To see that, we recall that the leading contribution to the transitions between two excited states  $|m\rangle$ and $|n\rangle$ is expressed via the matrix element:
 ${\cal F}_{mn} = \langle m|{\cal F}|n\rangle = {f}_{ij}({\cal X}^{m\ast}_{ik}{\cal X}^{n}_{jk} + {\cal Y}^{m\ast}_{jk}{\cal Y}^{n}_{ik})$  summed over the repeated quasiparticle state indices of both protons and neutrons. Here $\cal F$ is the transition operator which, in the case of electromagnetic transitions, has non-vanishing contributions ${f}_{ij}$ only for the proton components, except for the case of dipole transitions where subtraction of the translation motion induces non-zero neutron effective charges \cite{Suhonen}. Thus, the dipole states with dominant proton $2q$ components are much more likely to gamma decay to the variety of excited states, while the states associated with the neutron skin oscillation predominantly decay to the ground state. 

Figure \ref{fig5}, together with the matrix element ${\cal F}_{mn}$ expressed via recoupled transition densities, connects the purely theoretical information about the internal microscopic structure of the LEDS peaks with the spectroscopic observables. Although the particular details of the state composition can be basis-dependent, there are general features of the spectrum that are experimentally detectable. 
Identification of the group of intruder states with enhanced decay probability to excited states and thus separable from the remaining strength  
decaying primarily to the ground state further clarifies the two-component structure of the LEDS 
experimentally observed in the comparison of isoscalar and isovector probes.
Preferential g.s.\ decay of the lower-energy component in $^{120}$Sn has been independently demonstrated in Ref.~\cite{Weinert2021}.

\begin{figure}
\vspace{-1.5cm}
\includegraphics[width=\columnwidth]{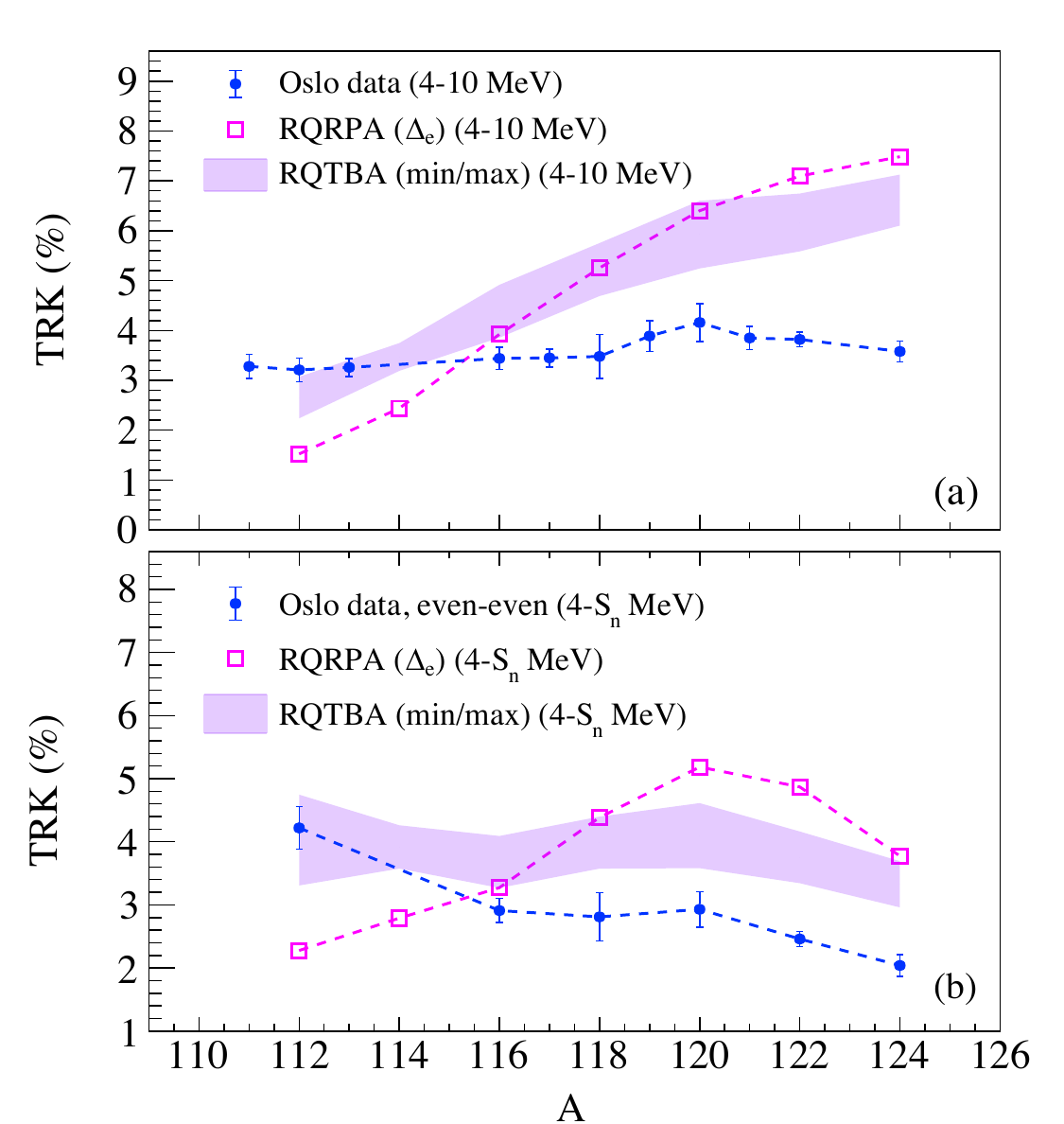}
\caption{\label{fig6}
Evolution of the experimental LEDS in the energy range (a) $4 -10$ MeV and (b) from $4$ MeV to the neutron threshold $S_n$  for even-mass Sn isotopes with mass numbers $A = 112 -124$ compared with predictions of the RQRPA (open squares) and RQTBA  (purple band, spanning the uncertainties due to the poorly constrained pairing interaction strength using $\Delta_e$, $\Delta_e (2^+_1)$, and three-point $\Delta_3$ adjusted to the odd-even mass staggering) calculations.
The dashed lines are to guide the eye.}
\end{figure}

The performed calculations and their analysis indicate that the RQTBA description, although significantly improved compared to RQRPA and helpful in interpreting the two-component structure of LEDS, calls for further refinements to achieve spectroscopic accuracy.  The continuum contribution, although small in the IVDGR region, is important in the reproduction of the Oslo data, and the thermal continuum is an adequate concept here \cite{LitvinovaBelov2013}.
Moreover, the continuum may play a decisive role around the particle emission threshold and intervene in the analysis employed for separating PDR and IVGDR. This is illustrated in Fig.~\ref{fig6}, where the integral experimental and theoretical strengths as functions of mass number are compared for the energy intervals 4 $\leq E \leq E_{max}$ with $E_{max} = 10$ MeV (a) and $E_{max} = S_n$ (b).
For $E_{max} = 10$ MeV, both model results show a monotonous increase between mass numbers 112 and 124 from about 1.5\% to 7.5\% of the TRK sum rule for RQRPA ($\Delta_e$) and 2.5\% to 6.5\% for RQTBA ($\Delta_e$), respectively.
The purple band of the RQTBA results reflects the uncertainties due to the poorly constrained pairing interaction strength calculated with the two approaches discussed above and the three-point ($\Delta_3$) values adjusted to the odd-even mass staggering.
In contrast, the experimental strengths are approximately constant with values $3.5 - 4$\%.
For the case of $^{120}$Sn, the models predict extra strength compared to the data in the energy region $7 - 10$  MeV, leading to a total value of 6\%. 
As seen in Fig.~\ref{fig6} (b), cutting out the contribution from  $S_n \leq E \leq 10$ MeV or adding those from $10 \leq E \leq S_n$ MeV may change the theoretical integral strength notably and even reverse its trend at large and small neutron numbers, respectively. 
A complete response theory, besides better constrained pairing correlations, should thereby take into account the continuum, including the multiparticle escape. Other many-body effects, such as a more complete set of phonons (in particular, those of unnatural parity and isospin-flip), complex ground-state correlations, and higher-complexity configurations are expected to generate the richer fine structure of the LEDS, especially at the lowest energies. 
 

\section{Summary and conclusions}

We present a systematic study of the low-energy electric dipole strength in Sn isotopes ranging from $A = 111$ to 124 based on the data for $^{111-113,116,118-122,124}$Sn obtained with the Oslo method and the study of $^{112,114,116,118,120,124}$Sn with relativistic Coulomb excitation in forward angle ($p,p^\prime$) scattering.  
The tin chain is of particular interest since the similarity of low-energy structure in the neighboring isotopes allows us to single out the impact of neutron excess on the evolution of the LEDS.
The combined data cover an energy range of $2 - 20$ MeV which permits a decomposition into the contribution from the low-energy tail of the IVGDR and possible resonance-like structures with minimal assumptions.
One finds in all cases a resonance at about 8.3 MeV, exhausting an approximately constant fraction of the TRK sum rule with a local maximum at $^{120}$Sn which might be attributed to shell structure effects.

For the isotopes with $A \geq 118$, a consistent description of the data suggests introducing a second resonance centered at about 6.5 MeV. 
The comparison with the results from the isoscalar probes and the $(\gamma,\gamma^\prime)$ reaction indicates that it might represent the IV response of the PDR. 
Its strength corresponds to a relatively small fraction of the total LEDS and demonstrates an approximately linear dependence on mass number.
Because of the schematic decomposition,
the absolute values might have some model dependence. 
Nevertheless, these results demonstrate that taking the full LEDS as representative of the PDR strength (as done in many discussions of a possible relation to the neutron skin thickness) is clearly unfounded.
The small PDR strength also challenges its interpretation of being due to neutron skin oscillations, which require some degree of collectivity.
However, this question cannot be settled based on the IV response alone.

Self-consistent microscopic calculations rooted in ab initio EOM theory and employing effective meson-exchange interaction can reproduce the experimental features of LEDS reasonably well provided qPVC is included.
Consideration of qPVC improves the description essentially by generating the necessary richness of the spectrum, in contrast to simplistic QRPA confined by $2q$ configurations forming LEDS dominated by one peak (cf.~Ref.~\cite{Markova2024}). In this work, leading-order qPVC was included in RQTBA, which allowed for decomposing theoretical LEDS into distinct structures identified in the experiment. A state-by-state analysis of the transition densities enabled the identification of states above 8 MeV whose microscopic composition is essentially different from those fragmented from primordial RQRPA states, namely, proton contribution prevails in their $2q$ content. We argue that -- in accordance with experimental observations -- the upper part of the LEDS dominated by such states has an enhanced probability of cascade decay, while its lower-energy counterpart, representing predominantly neutron oscillations, decays mostly to the ground state. Further analysis showed that the latter part of the strength is likely to be associated with a single resonance-like structure in agreement with experimental data, while the upper part of the LEDS is more fragmented in heavier isotopes and interferes with the low-energy shoulder of IVGDR.
Clarifying this interference, as well as obtaining more accurate spectroscopic results, calls for furthering theoretical effort in both the interaction and many-body aspects. The most pressing issues obscuring refined computation remain in the segments of pairing correlations, continuum, and complex configurations beyond the leading-order qPVC including those of the nuclear ground state.

Further experimental work is required to elucidate the true nature of the PDR.
One way is systematic studies of the IV and IS response along isotopic or isotonic chains, similar to the one presented here.
For the Sn case, experiments at RIB facilities in inverse kinematics, analogous to the pioneering study of E1 strength in $^{130,132}$Sn \cite{Adrich2005}, but including the information on $\gamma$ decay \cite{Wieland2009,Rossi2013} to extract the GSF below neutron thresholds, would be of particular interest.
New experimental observables like the particle-hole structure investigated in transfer reactions \cite{Weinert2021,Spieker2020} or transition current densities from transverse electron scattering \cite{Neumann-Cosel2023} may help to clarify the underlying structure.


\section*{Acknowledgements}
This work was supported in part by the National Science Foundation under Grant No.\ OISE-1927130 (IReNA) and by the Norwegian Research Council Grants 325714 and 263030. 
P.v.N.-C.\ acknowledges support by the Deutsche Forschungsgemeinschaft (DFG, German Research Foundation) under Grant No.\ SFB 1245 (Project ID 279384907). 
The work of E.L.\ was partly performed at the Aspen Center for Physics supported by the National Science Foundation (NSF) grant PHY-2210452 and 
partly supported by the GANIL Visitor Program, US-NSF grant PHY-2209376, and US-NSF Career grant PHY-1654379. 

\section*{Appendix A. Supplementary material}

Supplementary material related to this article is provided below. 



\bibliographystyle{elsarticle-num} 
\bibliography{tin_2023_plb_new}






%

\afterpage{\blankpage}

\includepdf[pages=-]{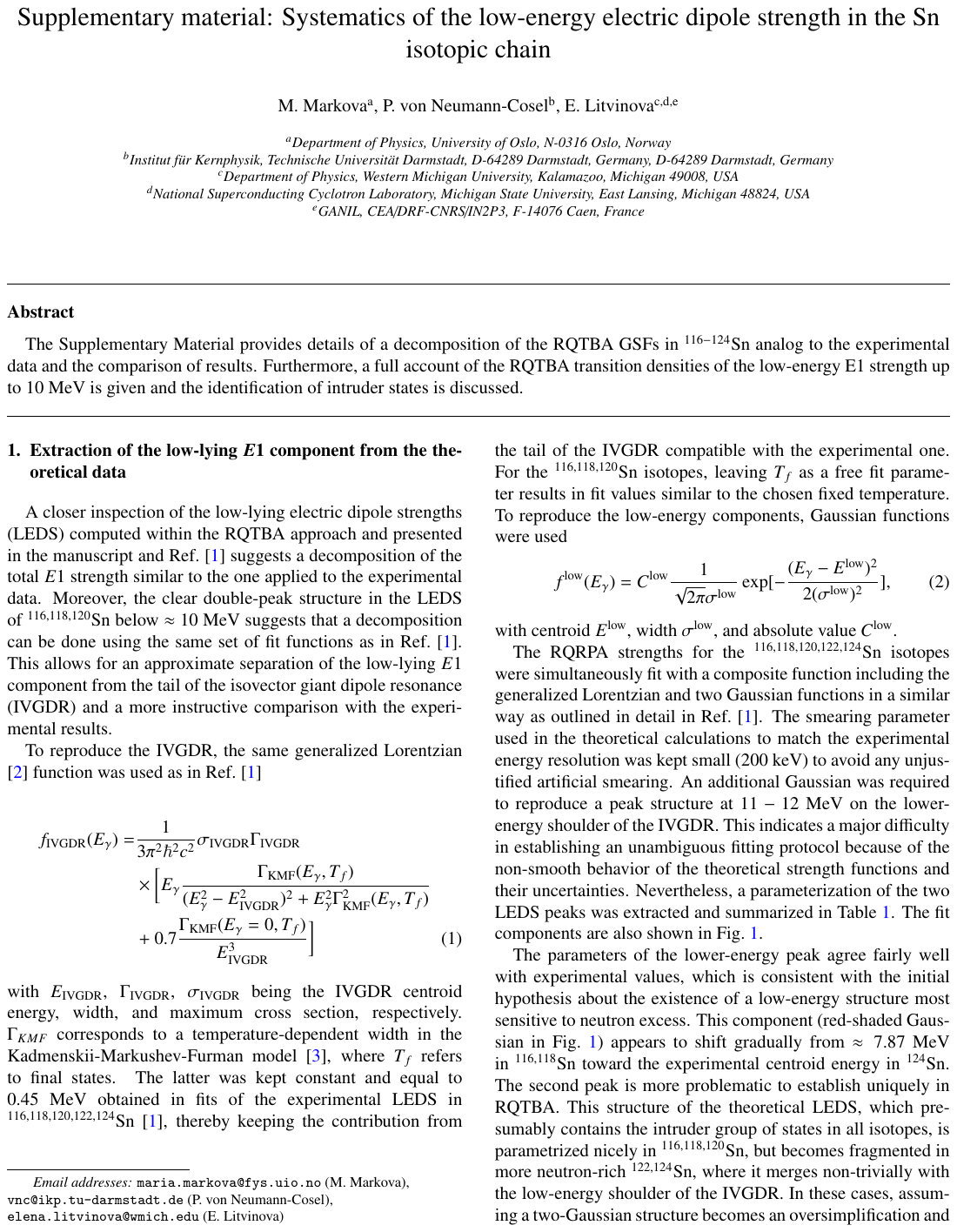}
\end{document}